\documentstyle[aps,epsf,twocolumn]{revtex}

\begin{document}
 \makeatletter
 \def\preprint#1{%
    \def\@preprint{\noindent\hfill\hbox{#1}\vskip 10pt}%
 }
\preprint{\begin{tabular}{l}
      hep-ph/0106192 \\
      CPHT/S029.0601 \\
      DESY-01-082
    \end{tabular}
 }
%
%
\title{Generalized parton distributions in the deuteron}
\author{E. R. Berger$^1$, F. Cano$^2$, M. Diehl$^3$ and B. Pire$^1$}
\address{$^1$CPhT, {\'E}cole Polytechnique, F-91128 Palaiseau, France \\
$^2$ DAPNIA/SPhN, CEA/Saclay, F-91191 Gif sur Yvette Cedex, France \\
$^3$Deutsches Elektronen-Synchroton DESY,
 D-22603 Hamburg, Germany}
%
%
\maketitle
\begin{abstract}
{\it Abstract.} We introduce generalized quark and gluon distributions in the
deuteron, which can be measured in exclusive processes like deeply
virtual Compton scattering and meson electroproduction. We discuss the
basic properties of these distributions, and point out how they probe
the interplay of nucleon and parton degrees of freedom in the
deuteron wave function.
\end{abstract}
%
\narrowtext
\noindent
{\it Introduction.}  The partonic structure of the deuteron has been
explored in terms of the parton distributions accessible in deep
inelastic scattering \cite{Hoodbhoy:1989am}, and in terms of the form
factors measured in elastic lepton-deuteron processes
\cite{Brodsky:1983vf,Garcon:2001sz}. It is natural to ask what can be
learned from generalized parton distributions (GPDs), introduced not
long ago in \cite{Muller:1994fv,Radyushkin:1997ki}. For the nucleon
case it has been shown that these quantities contain unique
information about the dynamics of quarks and gluons in QCD bound
states, beyond what can be unraveled from ordinary parton
distributions and form factors.
Here we extend these studies to the case of the deuteron, with the aim
of providing the theoretical framework to analyze and interpret
present and future measurements with deuteron targets. We restrict
ourselves to parton distributions of twist two, and to the parton
helicity conserving sector, which is relevant in most phenomenological
applications. Quark and gluon helicity flip GPDs can be treated with
the same methods.
\medskip

\noindent
{\it Generalized quark distributions for the deuteron.}  As in the
nucleon case, the GPDs for the deuteron are defined through
non-diagonal matrix elements of quark-antiquark operators on the light
cone. Their general decomposition can be written as
\begin{eqnarray}
V_{\lambda'\lambda} &=&
  \int \frac{d \kappa}{2 \pi}\,
  e^{i x \kappa P.n}
  \langle p', \lambda' |\,
  \bar{\psi}(-\kappa  n)\, \gamma.n\, \psi(\kappa n)
  \,| p, \lambda \rangle
\nonumber \\
 & = & \sum_{i}
  \epsilon'^{\ast \beta}  V^{(i)}_{\beta \alpha}\,
  \epsilon^{\alpha}\, H_{i}(x,\xi,t) ,
\label{vaten}
\\
A_{\lambda'\lambda} &=&
  \int \frac{d \kappa}{2 \pi}\,
  e^{i x \kappa P.n}
  \langle p', \lambda' |\,
  \bar{\psi}(-\kappa n)\, \gamma.n \gamma_5\, \psi(\kappa n)
  \,| p, \lambda \rangle
\nonumber \\
& = & \sum_{i}
  \epsilon'^{\ast \beta}  A^{(i)}_{\beta \alpha}\,
  \epsilon^{\alpha}\, \tilde{H}_{i}(x,\xi,t) ,
\label{axvaten}
\end{eqnarray}
where $n$ is a light-like four-vector, $n^2=0$. The incoming and
outgoing deuterons respectively have momenta $p$, $p'$, helicities
$\lambda$, $\lambda'$, and polarization vectors $\epsilon=\epsilon
(p,\lambda)$ and $\epsilon'=\epsilon'(p',\lambda' )$.  We write $P =
p+p'$ and $\Delta = p' - p$, and choose Ji's variables $x$, $\xi =
-(\Delta.n) /(P.n)$, $t= \Delta^2$ as arguments of the GPDs $H_i$ and
$\tilde{H}_i$.
The tensors $V^{(i)}$ and $A^{(i)}$ depend on the four-vectors $p$,
$p'$, and $n$. One need only keep tensors which do not vanish when
contracted with $\epsilon_{\alpha}^{\phantom{'}}$,
$\epsilon'_{\beta}$, given the orthogonality conditions $\epsilon.p =
\epsilon'.p' = 0$ of the polarization vectors. With the constraints
from parity invariance we find that the $V^{(i)}_{\beta \alpha}$ are
linear combinations of the five tensor structures
\begin{equation}
  \label{vecten}
\{ g_{\beta \alpha},\hspace{0.5em}
   p_{\beta} n_{\alpha} ,\hspace{0.5em}
   n_{\beta} p'_{\alpha} ,\hspace{0.5em}
   p_{\beta} p'_{\alpha} ,\hspace{0.5em}
   n_{\beta} n_{\alpha} \}.
\end{equation}
Similarly, the $A^{(i)}_{\beta \alpha}$ are linear combinations of the
seven tensors
\begin{eqnarray}
  \label{axten}
 & & \{  \epsilon_{\mu \nu \beta \alpha}\,
            p^{\mu}p'^{\nu},\hspace{0.5em}
  \epsilon_{\mu \nu \beta \alpha}\, n^{\mu}p^{\nu},\hspace{0.5em}
  \epsilon_{\mu \nu \beta \alpha}\, n^{\mu}p'^{\nu},  \nonumber \\
 & & \hspace{0.5em}
   \epsilon_{\mu \nu \rho \beta}\,
        p^{\mu}p'^{\nu} n^{\rho} n_{\alpha} ,\hspace{0.5em}
   \epsilon_{\mu \nu \rho \beta}\,
        p^{\mu}p'^{\nu} n^{\rho} p'_{\alpha} ,
\nonumber \\
 & & \hspace{0.5em}
   \epsilon_{\mu \nu \rho \alpha}\,
        p^{\mu}p'^{\nu} n^{\rho} n_{\beta} ,\hspace{0.5em}
   \epsilon_{\mu \nu \rho \alpha}\,
        p^{\mu}p'^{\nu} n^{\rho} p_{\beta}
 \} .
\end{eqnarray}
Using the Schouten identities \cite{deWit:1986} one can show that only
four out of these seven are linearly independent.
The first $x$ moments of generalized parton distributions are elastic
form factors. To keep the corresponding relations simple we take among
the tensors $V^{(i)}$, $A^{(i)}$ those which appear in the
conventional form factor decomposition of the vector and axial
currents \cite{Frederico:1992vm}:
\begin{eqnarray}
\label{local}
\lefteqn{
  \langle p', \lambda' |\,
    \bar{\psi}(0)\, \gamma^\mu\, \psi(0) \,| p, \lambda \rangle
  = {}- G_1(t)\, (\epsilon'^*.\epsilon) P^\mu
\rule[-2ex]{0pt}{2ex}
} \hspace{2em}  \nonumber \\
&+& G_2(t) \left[\epsilon^\mu (\epsilon'^*.P)
      + \epsilon'^{* \mu} (\epsilon.P)\right]
\nonumber \\
&-& G_3(t)\, (\epsilon .P)(\epsilon'^* .P)\,
\frac{P^\mu}{2 M^2} ,
\nonumber \\
\lefteqn{
  \langle p', \lambda' |\,
    \bar{\psi}(0)\, \gamma^\mu \gamma_5\, \psi(0) \,| p, \lambda \rangle
  = {}- i \tilde{G}_1(t)\, \epsilon^\mu{}_{\!\alpha \beta \gamma}\,
    \epsilon'^{* \alpha} \epsilon^\beta P^\gamma
\rule[-1ex]{0pt}{4ex}
} \hspace{2em} \nonumber  \\
&+& i \tilde{G}_2(t)\,
          \epsilon^\mu{}_{\!\alpha \beta \gamma}\,
    \Delta^\alpha P^\beta\, \frac{\epsilon^\gamma (\epsilon'^*.P)
    + \epsilon'^{* \gamma} (\epsilon .P)}{M^2} ,
\end{eqnarray}
where our convention for the antisymmetric tensor is $\epsilon_{0123}
= 1$ and M is the deuteron mass.  The matrix elements are here defined
flavor by
flavor; to get the conventional form factors, one must weight with
electromagnetic or weak charges and sum over flavors. For the matrix
elements of
the non-local operators we define
\begin{eqnarray}
\lefteqn{
V_{\lambda'\lambda} =
{}- (\epsilon'^* .\epsilon)\, H_1
  + \frac{(\epsilon .n) (\epsilon'^* .P)
  + (\epsilon'^* .n) (\epsilon .P)}{P.n}\, H_2
} \hspace{1em}  \nonumber\\
&& {}- \frac{(\epsilon .P)(\epsilon'^* .P)}{2 M^2}\, H_3
 + \frac{(\epsilon .n) (\epsilon'^* .P)
  - (\epsilon'^* .n) (\epsilon .P)}{P.n}\, H_4
\nonumber \\
&& {}+ \Big\{
4 M^2\, \frac{(\epsilon .n)(\epsilon'^* .n)}{(P.n)^2}
 +\frac{1}{3} (\epsilon'^* .\epsilon) \Big\}
 H_5 \; ,
\nonumber \\
\lefteqn{
A_{\lambda'\lambda} =
{}- i \frac{\epsilon_{\mu \alpha \beta \gamma}
   n^\mu \epsilon'^{*\, \alpha}
  \epsilon^\beta P^\gamma}{P.n}\,
\tilde{H}_1
} \hspace{1em}  \nonumber \\
&& {}+ i \frac{\epsilon_{\mu \alpha \beta \gamma}\, n^\mu
 \Delta^\alpha P^\beta}{P.n}\,
 \frac{ \epsilon^\gamma (\epsilon'^*.P) +
        \epsilon'^{* \,\gamma} (\epsilon .P) }{M^2}\,
\tilde{H}_2
\nonumber \\ &&
+ i \frac{\epsilon_{\mu \alpha \beta \gamma}\, n^\mu
 \Delta^\alpha P^\beta}{P.n}\,
 \frac{ \epsilon^\gamma (\epsilon'^*.P) -
        \epsilon'^{* \,\gamma} (\epsilon .P) }{M^2}\,
\tilde{H}_3
\nonumber \\ & &
{}+ i \frac{\epsilon_{\mu \alpha \beta \gamma}\, n^\mu
 \Delta^\alpha P^\beta}{P.n}\,
 \frac{ \epsilon^\gamma (\epsilon'^*.n) +
        \epsilon'^{* \,\gamma} (\epsilon .n) }{P.n}\,
\tilde{H}_4\; .
\label{final}
\end{eqnarray}
Since it is determined by the quark operators in Eqs.~(\ref{vaten})
and (\ref{axvaten}), the $Q^2$ evolution of the GPDs $H_i$ and
$\tilde{H}_i$ is exactly the same as for spin $1/2$ targets, worked
out in
\cite{Muller:1994fv,Radyushkin:1997ki,Blumlein:1999sc,Belitsky:2000hf}.
\medskip

\noindent
{\it Time reversal properties and sum rules.}  The action of the time
reversal operator on the matrix element (\ref{vaten}) leads to the
relation
\begin{eqnarray}
&& \epsilon'^{\ast \beta}
\epsilon^{\alpha} V^{(i)}_{\beta \alpha}\,
(P,\Delta,n)\, H_{i}(x,\xi,t)
\nonumber \\
&=& \epsilon^{\ast \beta}
\epsilon'^{\alpha} V^{(i)}_{\beta \alpha}\,
(P,-\Delta,n)\, H_{i}(x,-\xi,t) ,
\label{timereversal}
\end{eqnarray}
where we have made explicit the dependence of the tensors $V^{(i)}$ on
the relevant four-vectors. Taking the complex conjugate of
Eq. (\ref{vaten}) we get
\begin{eqnarray}
&& \epsilon'^{ \beta}
\epsilon^{\ast \alpha}\, V^{\ast (i)}_{\beta \alpha} (P,\Delta,n)\,
H_{i}^\ast (x,\xi,t) \nonumber \\
&=& \epsilon^{\ast \beta}
\epsilon'^{\alpha}\, V^{(i)}_{\beta \alpha}(P,-\Delta,n)\,
H_{i}(x,-\xi,t) .
\label{conjugate}
\end{eqnarray}
For the axial vector case, we obtain relations analogous to
(\ref{timereversal}) and (\ref{conjugate}) by replacing $V^{(i)}$ with
$A^{(i)}$ and $H_i$ with $\tilde{H}_i$. Combining these conditions we
find that all nine GPDs are real. However, their behavior under time
reversal is not uniform and we have:
\begin{eqnarray}
H_i(x,\xi,t) &  = & H_i(x,-\xi,t) \hspace{2em} (i=1,2,3,5) ,
\nonumber \\
H_4(x,\xi,t) &  = & - H_4(x,-\xi,t) , \hspace{2em}
\nonumber \\
\tilde{H}_i(x,\xi,t) & = & \tilde{H}_i(x,-\xi,t) \hspace{2em}
(i=1,2,4) ,
\nonumber \\
\tilde{H}_3(x,\xi,t) & = & - \tilde{H}_3(x,-\xi,t) .
\label{time}
\end{eqnarray}
Note that in the non-forward case, time reversal invariance fixes the
phase of the generalized parton distributions and determines their behavior
under sign change of the skewedness parameter $\xi$, but does
\emph{not} limit the number of GPDs
\cite{Diehl:2001pm}.
Integrating $V_{\lambda'\lambda}$ and $A_{\lambda'\lambda}$ over $x$
one obtains the local matrix elements (\ref{local}) contracted with
$n^\mu /(P.n)$. Since the tensors that accompany the distributions
$H_i$, $\tilde{H}_i$ in Eq.~(\ref{final}) are linearly independent, we
obtain the sum rules
\begin{eqnarray}
\int_{-1}^1 dx H_i(x,\xi,t)&  = & G_i(t) \hspace{3em} (i=1,2,3) ,
\nonumber \\
\int_{-1}^1 dx \tilde{H}_i(x,\xi,t) & = &  \tilde{G}_i(t) \hspace{3em}
(i=1,2) ,
\nonumber \\
\int_{-1}^1 dx H_4(x,\xi,t) &=& \int_{-1}^1 dx \tilde{H}_3(x,\xi,t)
\;=\; 0 ,
\nonumber \\
\int_{-1}^1 dx H_5(x,\xi,t) &=& \int_{-1}^1 dx \tilde{H}_4(x,\xi,t)
\;=\; 0 .
\label{h5moment}
\end{eqnarray}
The integrals over $H_4$, $\tilde{H}_3$ and $H_5$, $\tilde{H}_4$ do
not correspond to form factors of the local vector or axial
currents and therefore vanish. In the case of $H_4$ and $\tilde{H}_3$
this is due to time reversal constraints, whereas the definitions of
$H_5$ and $\tilde{H}_4$ involve the tensor $n^\mu n^\nu / (P.n)^2$,
whose analog cannot appear in the decomposition of the local currents
due to Lorentz invariance.
\medskip

\noindent
{\it The forward limit.}  Let now study the forward limit of the GPDs,
which defines the usual parton distributions.  In the parton model,
i.e., at leading twist and leading order in $\alpha_s$ there are three
independent structure functions in deep inelastic scattering, $F_1$,
$b_1$, $g_1$, whose probabilistic interpretation in terms of quark
densities reads \cite{Hoodbhoy:1989am}
\begin{eqnarray}
F_1(x) & = & \frac{1}{2} \sum_q e_q^2 \frac{q^1 (x)
             + q^{-1} (x) + q^0 (x)}{3}
             + \{q \to \bar{q}\} , \nonumber \\
b_1(x) & = & \frac{1}{2} \sum_q e_q^2 \Big[ q^0 (x)
             - \frac{q^{1} (x) + q^{-1} (x)}{2} \Big]
             + \{q \to \bar{q}\} , \nonumber \\
g_1(x) & = & \frac{1}{2} \sum_q e_q^2 \Big[
             q^1_\uparrow (x)  - q^{-1}_\uparrow (x) \Big]
             + \{q \to \bar{q}\} .
\label{parton-model}
\end{eqnarray}
Here $q_{\uparrow (\downarrow)}^\lambda (x)$ represents the
probability to find a quark with momentum fraction $x$ and positive
(negative) helicity in a deuteron target of helicity
$\lambda$. The unpolarized quark densities $q^\lambda$ are defined as
$\smash{ q^{\lambda}(x) = q_{\uparrow}^\lambda(x) +
q_\downarrow^\lambda(x) }$. From parity one has $\smash{
q^\lambda_{\uparrow} = q^{-\lambda}_{\downarrow} }$. The densities for
antiquarks are defined in analogy. Note that the probabilistic
interpretation for $F_1$ and $g_1$ is similar to the one in the spin
$1/2$ case, whereas the function $b_1$ does not appear for spin $1/2$
targets.
In the forward limit the only structures in Eq.~(\ref{final}) that
survive are those proportional to $H_1$, $H_5$ and $\tilde{H}_1$,
because in that limit we have $\Delta=0$ and $\epsilon.P = \epsilon'.P
= 0$. Using the results for helicity amplitudes given below, one gets:
\begin{eqnarray}
\label{forward}
H_1(x,0,0)         & = &  \frac{q^1(x) + q^{-1}(x) + q^0(x)}{3} ,
\nonumber \\
H_5(x,0,0)         & = &  q^0(x) - \frac{q^{1}(x) + q^{-1}(x)}{2} ,
\nonumber \\
\tilde{H}_1(x,0,0) & = &  q^1_\uparrow (x)  - q^{-1}_\uparrow (x)
 \rule{0pt}{3ex}
\end{eqnarray}
for $x>0$. The corresponding relations for $x<0$ involve the antiquark
distributions at $-x$, with an overall minus sign in the expressions
for $H_1$ and $H_5$. With Eq.~(\ref{h5moment}) we thus have
\begin{eqnarray}
0 &=& \int_{-1}^1 dx H_5(x,0,0)
\nonumber \\
&=& \int_{0}^1 dx\,
 \Big[ q^0(x) - \frac{q^{1}(x) + q^{-1}(x)}{2} \big]
- \{q \to \bar{q}\} ,
\end{eqnarray}
and recover the parton model sum rule $\int_0^1 b_1(x)=0$ of
\cite{Close:1990zw}, which was obtained under the assumption that the
quark sea $q - \bar{q}$ does not contribute to this integral.
\medskip

\noindent
{\it Helicity amplitudes.} In the region $\xi<x<1$ our GPDs can be
represented in terms of amplitudes for the scattering of a quark on a
deuteron \cite{Diehl:2001pm}, defined as
\begin{equation}
{\cal A}_{\lambda'\pm, \lambda\pm} = \frac{1}{2} \,
   ( V_{\lambda'\lambda} \pm  A_{\lambda'\lambda} ).
\label{def-amp}
\end{equation}
with $\pm$ referring to the helicities of the quarks. With the
constraints
\begin{equation}
{\cal A}_{-\lambda'-\mu, -\lambda-\mu} =
(-1)^{\lambda'-\lambda} {\cal A}_{\lambda'\mu,\lambda\mu}
\label{P}
\end{equation}
from parity invariance there are nine independent quark helicity
conserving amplitudes. Since ${\cal A}_{0+,0+} = {\cal A}_{0-,0-}$ we
have only four quark helicity dependent distributions $\tilde{H}_i$,
compared with the five quark helicity independent $H_i$. Time reversal
invariance gives
\begin{equation}
{\cal A}(x,\xi,t)_{\lambda\mu,\lambda'\mu} =
(-1)^{\lambda'-\lambda} {\cal A}(x,-\xi,t)_{\lambda'\mu,\lambda\mu}
\; ,
\label{T}
\end{equation}
and thus does not further reduce the number of GPDs, as remarked
above. To define the polarization of the incoming deuteron we
introduce
\begin{eqnarray}
\epsilon^{(0) \mu} & = & \frac{1}{M}\,
  \Big( p^\mu-\frac{2 M^2}{1+\xi}\, \frac{n^\mu}{P.n} \Big) ,
\nonumber \\
\epsilon^{(1) \mu} & = & {}- \frac{1}{\sqrt{(1-\xi^2)(t_0-t)}}\,
  \Big( (1+\xi)\, p'^\mu - (1-\xi)\, p^\mu
\nonumber \\
& & \hspace{9em}
  {}- \frac{\xi (t_0-t) - t_0}{\xi}\, \frac{n^\mu}{P.n} \,\Big) ,
\nonumber \\
\epsilon^{(2) \mu} & = & \frac{1}{\sqrt{(1-\xi^2)(t_0-t)}}\,
  \frac{2 \epsilon^{\mu}{}_{\nu \alpha \beta}\,
  p'^\nu p^\alpha n^\beta}{P.n} ,
\label{pol-vectors}
\end{eqnarray}

where $t_0 = -4 M^2 \xi^2 /(1-\xi^2)$. The vectors $\epsilon(0) =
\epsilon^{(0)}$ and $\epsilon(\pm 1) = \mp (\epsilon^{(1)} \pm i
\epsilon^{(2)}) /\sqrt{2}$ then correspond to definite light-cone
helicity \cite{Brodsky:1998de}. This approximately coincides with
usual helicity in frames where the deuteron moves fast, provided that
$\mbox{sgn}(p^3) = - \mbox{sgn}(n^3)$.  The polarizations for the
outgoing deuteron are obtained from Eq.~(\ref{pol-vectors}) by the
exchange $p^\mu \leftrightarrow p'^\mu$, $\xi \leftrightarrow -\xi$
and an overall sign change for $\epsilon^{(1)}$ and
$\epsilon^{(2)}$. With this we get
\begin{eqnarray}
{\cal A} _{++,++} & = & \frac {H_1}{2}-\frac{H_5}{6} +
 \frac{D H_3}{2} + \frac{\tilde{H}_1}{2}  + 2 D (\tilde{H}_2 +
 \xi \tilde{H}_3) ,
\nonumber \\
{\cal A} _{0+,0+} &  = &  \frac{H_1}{2}
 -\xi H_4 + \frac{1}{3} \Big(1-\frac{3}{2} \xi^2\Big) H_5
\nonumber \\
 & - & \Big(D-\frac{\xi^2}{1-\xi^2}\Big) \Big(H_1 - H_2  -
 \xi H_4 - \frac{1}{3} H_5 \Big)
\nonumber \\
 & - & \Big( D^2-\frac{\xi^2}{(1-\xi^2)^2} \Big) H_3 ,
\nonumber \\
{\cal A} _{-+,++} & = &  - D\, \Big( \, \frac{1}{2} H_3
 + 2 (\xi \tilde{H}_2 + \tilde{H}_3) \Big) ,
\nonumber \\
{\cal A} _{0+,++}  & = & \sqrt{\frac{D (1-\xi)}{2 (1+\xi)}}\,
 \left[H_1 - \frac{1-\xi}{2} (H_2 -H_4) \right.
\nonumber \\
 & & \hspace{4em} \left.
 {}- \frac{1}{3} H_5
 + \Big(D - \frac{\xi}{1-\xi^2}\Big) H_3 \,\right]
\nonumber \\
 & + & \sqrt{2 D (1- \xi^2)}\,
 \left[ \frac{1}{4}(\tilde{H}_1 + (1-\xi) \tilde{H}_4) \right.
\nonumber \\
 & & \hspace{4em} \left.
 {}+ \Big(D - \frac{\xi}{1-\xi^2}\Big) (\tilde{H}_2
 + \tilde{H}_3) \,\right] ,
\label{helamp}
\end{eqnarray}
where $D = (t_0-t) /(4 M^2)$. The remaining amplitudes can be easily
obtained from the relations (\ref{P}), (\ref{T}) and
\begin{equation}
{\cal A}_{\lambda'-\mu, \lambda-\mu} =
{\cal A}_{\lambda'\mu,\lambda\mu} (\tilde{H}_i \longrightarrow
-\tilde{H}_i) .
\end{equation}
As required by angular momentum conservation, one gets a factor
$\sqrt{t_0-t}$ for each unit of helicity flip. Note that $H_{2,4}$ and
$\tilde{H}_4$ only appear with longitudinal deuteron polarization, and
that the only GPDs appearing in double helicity flip amplitudes are
$H_3$ and $\tilde{H}_{2,3}$. In the forward limit we have $\smash{
{\cal A}(x,0,0)_{\lambda+,\lambda+} = q_\uparrow^\lambda(x) }$,
$\smash{ {\cal A}(x,0,0)_{\lambda-,\lambda-} = q_\downarrow^\lambda(x)
}$ and find the relations (\ref{forward}).
\medskip

\noindent
{\it Gluon distributions.}  Let us turn to the gluon distributions in
the deuteron. Instead of the matrix elements (\ref{vaten}),
(\ref{axvaten}) we now have
\begin{eqnarray}
\lefteqn{
  4 \frac{n_\alpha n_\beta}{P.n}
  \int \frac{d \kappa}{2 \pi} e^{i x \kappa P.n}
  \langle p', \lambda ' |
  F^{\alpha\mu}(-\kappa n)\, F_{\mu}{}^\beta(\kappa n)
  | p, \lambda \rangle } \hspace{7em}
\nonumber \\
 & = & \sum_{i}
  \epsilon'^{\ast \beta} V^{(i)}_{\beta \alpha}\,
  \epsilon^{\alpha}\, H_{i}^g(x,\xi,t) ,
\nonumber \\
\lefteqn{ -4i \frac{n_\alpha n_\beta}{P.n}
  \int \frac{d \kappa}{2 \pi} e^{i x \kappa P.n}
  \langle p', \lambda ' |
  F^{\alpha\mu}(-\kappa n)\, \tilde{F}_{\mu}{}^\beta(\kappa n)
  | p, \lambda \rangle } \hspace{7em}
\nonumber \\
& = & \sum_{i}
  \epsilon'^{\ast \beta} A^{(i)}_{\beta \alpha}\,
  \epsilon^{\alpha}\, \tilde{H}^g_{i}(x,\xi,t) ,
\label{vaten-g}
\end{eqnarray}
with $\tilde{F}^{\alpha\beta}=\frac{1}{2}
\epsilon^{\alpha\beta\gamma\delta} F_{\gamma\delta}$.  We take the
same tensors $V^{(i)}$, $A^{(i)}$ as for quark distributions, given in
Eq.~(\ref{final}).  Note that the $H_i^g$ are even and the
$\tilde{H}_i^g$ odd in~$x$. Their behavior under time reversal is the
same as in (\ref{time}) for the corresponding quark distributions, and
the definitions and expressions of the helicity amplitudes ${\cal
A}^g_{\lambda'\mu, \lambda\mu}$ are also analogous to the quark case.
The forward limit is now $\smash{ {\cal A}^g(x,0,0)_{\lambda+,
\lambda+} = x g_{\uparrow}^\lambda (x) }$ and $\smash{ {\cal
A}^g(x,0,0)_{\lambda-, \lambda-} = x g_{\downarrow}^\lambda (x) }$,
with an extra factor $x$ compared to the quark case.
\medskip

\noindent
{\it Some Phenomenology.}  The deuteron GPDs can be accessed in hard
exclusive processes such as deeply virtual Compton scattering in $e d
\to e d \gamma$, and electroproduction $e d \to e d M$ of a meson
or a meson pair\cite{DGPT}. The relevant kinematical limit is that of large
invariant momentum transfer $Q^2$ to the lepton at fixed $x_B$ and $t$, where
the Bjorken variable $x_B$ is defined as in deep inelastic scattering.
Factorization formulae are the same as for nucleon targets
\cite{Radyushkin:1997ki,Collins:1997fb}, with the appropriate
replacement of the hadronic matrix elements, but the same hard
scattering kernels.  The $Q^2$ behavior of the amplitudes and the
selection rules for photon and meson helicities also remain the same,
since they depend on the hard-scattering process, not on the target
spin.
Electroproduction of a pseudoscalar meson selects the GPDs $\tilde
H_i$, vector meson production involves the $H^{\phantom{g}}_i$ and $H_i^g$, and
all distributions appear in Compton scattering. Notice that the isosinglet
nature of the deuteron simplifies the flavor structure of the GPDs and
thus of the scattering amplitudes. One consequence is that pion
exchange does not contribute to any of these processes, in contrast to
the nucleon case, where it may give important contributions through
the quark distribution $\tilde{E}$ \cite{Goeke:2001tz}.
In kinematics where the Bethe-Heitler process dominates in
electroproduction $e d \to e d \gamma$, one can use the methods of
\cite{Diehl:1997bu} to study Compton scattering through the
interference of the two processes. At sufficiently large $Q^2$ this
interference term gives access to a linear combination of GPDs,
weighted with the electromagnetic deuteron form factors
$G_{1,2,3}(t)$.
\medskip

\noindent
{\it Discussion.}  To get a feeling for the physics of GPDs in the
deuteron, consider the approximation where they are written as a
convolution of the nucleon GPDs with the light-cone wave function
$\psi_{p+n}$ for a proton and a neutron in the deuteron
\cite{Frankfurt:1981mk}. The struck nucleon then has to absorb the
entire momentum transfer, in particular its plus-component
parameterized by $\xi$. Assuming for simplicity that $\psi_{p+n}$ is
only nonzero if the plus-momentum fraction of the proton in the
deuteron is between $\smash{\frac{1}{2}} (1-w)$ and
$\smash{\frac{1}{2}} (1+w)$, one finds that all $H_i$ and
$\tilde{H}_i$ vanish for $\xi>w$. The $\xi$ dependence of the deuteron
GPDs thus reflects the width of the wave function $\psi_{p+n}$ in
longitudinal momentum fraction. For $\xi$ well above $w$ they provide
access to deuteron wave function components that cannot be described
in terms of individual nucleons. To unravel such components in
inclusive deep inelastic scattering at $x_B > 1$ has turned out to be
difficult. Approaching this region from below, the struck quark has to
take the entire momentum of a nucleon in the convolution picture, so
that not only the deuteron wave function $\psi_{p+n}$ becomes small
but also the parton density in the nucleon. This is not the case for
GPDs with their independent momentum variables $x$ and $\xi$\,: at
$\xi>w$ one can still have any value for the plus-momentum fraction of
the struck parton in the target.

We also note that, since in a convolution model $b_1$ requires a $d$
wave component in $\psi_{p+n}$ \cite{Hoodbhoy:1989am}, the same holds
for $H_5$. Double helicity flip amplitudes also need a $d$ wave
admixture: the helicity flip of a nucleon cannot exceed one unit, so
that orbital angular momentum is necessary. Hence $H_3$ and
$\tilde{H}_{2,3}$ involve the $d$ wave part of $\psi_{p+n}$.

In this letter we have focused on the case where the deuteron scatters
elastically. One may extend our study to the case where it dissociates
into a proton and neutron, or a more complicated hadronic system,
introducing appropriate transition GPDs. Factorization still holds in
the processes discussed above, provided that the invariant mass of the
dissociative system is small compared with the hard scale $Q^2$.

Let us finally stress that for low enough $\xi$ the generalized parton
distributions for elastic deuteron transitions are by no means
small. Neither is there any suppressing factor in the cross
sections. It should thus be possible to perform
exclusive electroproduction experiments on a deuteron target where
those on a nucleon are possible. We  hope for a rich harvest of
physics on this topic in the forthcoming years at existing facilities
such as DESY and Jefferson Lab.
\medskip

\noindent
{\it Acknowledgments.}  We acknowledge discussions with
J.~P.~Ralston and O.~V.~Teryaev.  This work is supported in part by
the TMR and IHP programmes of the European Union, Contracts
No.~HPRN-CT-2000-00130 and No.~FMRX-CT98-0194.  CPhT is UMR 7644 of
CNRS.



\begin{references}
\bibitem{Hoodbhoy:1989am}
P.~Hoodbhoy, R.~L.~Jaffe, A.~Manohar,
Nucl.\ Phys.\ B {\bf 312}, 571 (1989).
\bibitem{Brodsky:1983vf}
S.~J.~Brodsky, C.~Ji, G.~P.~Lepage,
Phys.\ Rev.\ Lett.\  {\bf 51}, 83 (1983).
\bibitem{Garcon:2001sz}
M.~Gar\c{c}on, J.~W.~Van Orden,
nucl-th/0102049.
\bibitem{Muller:1994fv}
D.~M{\"u}ller {\it et al.},
Fortsch.\ Phys.\  {\bf 42},  101 (1994)
[hep-ph/9812448];\\
%
X.~Ji,
Phys.\ Rev.\ Lett.\  {\bf 78}, 610 (1997)
[hep-ph/9603249].
\bibitem{Radyushkin:1997ki}
A.~V.~Radyushkin,
Phys.\ Rev.\ D {\bf 56}, 5524 (1997)
[hep-ph/9704207].
\bibitem{deWit:1986}
B.~de~Wit, J.~Smith, \emph{Field Theory in Particle Physics},
Vol.~I, Amsterdam (North Holland) 1986.
\bibitem{Frederico:1992vm}
T.~Frederico {\it et al.},
Phys.\ Rev.\ C {\bf 46}, 347 (1992);\\
%
R.~G.~Arnold, C.~E.~Carlson, F.~Gross,
Phys.\ Rev.\ C {\bf 23}, 363 (1981).
\bibitem{Blumlein:1999sc}
J.~Bl{\"u}mlein, B.~Geyer, D.~Robaschik,
Nucl.\ Phys.\ {\bf B560}, 283 (1999)
[hep-ph/9903520].
\bibitem{Belitsky:2000hf}
A.~V.~Belitsky, A.~Freund, D.~M\"uller,
Nucl.\ Phys.\  {\bf B574}, 347 (2000)
[hep-ph/9912379].
\bibitem{Diehl:2001pm}
M.~Diehl,
Eur.\ Phys.\ J.\ C {\bf 19}, 485 (2001).
\bibitem{Close:1990zw}
F.~E.~Close, S.~Kumano,
Phys.\ Rev.\ D {\bf 42}, 2377 (1990).
\bibitem{Brodsky:1998de}
S.~J.~Brodsky, H.~Pauli, S.~S.~Pinsky,
Phys.\ Rept.\  {\bf 301}, 299 (1998).
\bibitem{Collins:1997fb}
J.~C.~Collins, L.~Frankfurt, M.~Strikman,
Phys.\ Rev.\ D {\bf 56}, 2982 (1997)
[hep-ph/9611433];\\
%
J.~C.~Collins, A.~Freund,
Phys.\ Rev.\ D {\bf 59}, 074009 (1999)
[hep-ph/9801262];\\
%
X.~Ji, J.~Osborne,
Phys.\ Rev.\ D {\bf 58}, 094018 (1998)
[hep-ph/9801260].
\bibitem{Goeke:2001tz}
K.~Goeke, M.~V.~Polyakov, M.~Vanderhaeghen,
hep-ph/0106012.
\bibitem{Diehl:1997bu}
M.~Diehl, T.~Gousset, B.~Pire, J.~P.~Ralston,
Phys.\ Lett.\ B {\bf 411}, 193 (1997)
[hep-ph/9706344];\\
%
A.~V.~Belitsky, D.~M\"uller, L.~Niedermeier, A.~Sch\"afer,
Nucl.\ Phys.\ B {\bf 593}, 289 (2001)
[hep-ph/0004059].
%
\bibitem{DGPT}
M.~Diehl {\it et al.},
Phys. Rev. Lett. {\bf 81}, 1782  (1998).
%
\bibitem{Frankfurt:1981mk}
L.~L.~Frankfurt, M.~I.~Strikman,
Phys.\ Rept.\  {\bf 76}, 215 (1981);\\
%
J.~Carbonell {\it et al.},
Phys.\ Rept.\  {\bf 300}, 215 (1998)
[nucl-th/9804029];\\
%
G.~A.~Miller,
Prog.\ Part.\ Nucl.\ Phys.\  {\bf 45}, 83 (2000)
[nucl-th/0002059].
\end{references}
\end{document}